\documentclass[aps,prl,twocolumn,superscriptaddress]{revtex4-2}
\usepackage{graphicx}
\usepackage{dcolumn}
\usepackage{bm}
\usepackage{color}
\usepackage{lipsum}
\usepackage{lineno}
\usepackage{subfigure}
\usepackage{booktabs}
\usepackage{multirow}
\usepackage{tabu}
\usepackage{ltxtable, filecontents}

\usepackage{lipsum}
\usepackage{adjustbox}
\usepackage[colorlinks,urlcolor=blue,linkcolor=blue,citecolor=blue]{hyperref}
\usepackage{siunitx}
\begin{document}

\title{Charge equilibration of Laser-accelerated Carbon Ions in Foam Target}%

\author{Bubo Ma} \thanks{These authors have contributed equally to this work.}
\affiliation{MOE Key Laboratory for Nonequilibrium Synthesis and Modulation of Condensed Matter, School of Science,
Xi'an Jiaotong University, Xi'an 710049, China}
\author{Jieru Ren} \thanks{These authors have contributed equally to this work.}
\affiliation{MOE Key Laboratory for Nonequilibrium Synthesis and Modulation of Condensed Matter, School of Science,
Xi'an Jiaotong University, Xi'an 710049, China}
\author{Lirong Liu}
\affiliation{MOE Key Laboratory for Nonequilibrium Synthesis and Modulation of Condensed Matter, School of Science,
Xi'an Jiaotong University, Xi'an 710049, China}
\author{Wenqing Wei}
\affiliation{MOE Key Laboratory for Nonequilibrium Synthesis and Modulation of Condensed Matter, School of Science,
Xi'an Jiaotong University, Xi'an 710049, China}
\author{Benzheng Chen}
\affiliation{MOE Key Laboratory for Nonequilibrium Synthesis and Modulation of Condensed Matter, School of Science,
Xi'an Jiaotong University, Xi'an 710049, China}
\author{Shizheng Zhang}
\affiliation{MOE Key Laboratory for Nonequilibrium Synthesis and Modulation of Condensed Matter, School of Science,
Xi'an Jiaotong University, Xi'an 710049, China}
\author{Hao Xu}
\affiliation{MOE Key Laboratory for Nonequilibrium Synthesis and Modulation of Condensed Matter, School of Science,
Xi'an Jiaotong University, Xi'an 710049, China}
\author{Zhongmin Hu}
\affiliation{MOE Key Laboratory for Nonequilibrium Synthesis and Modulation of Condensed Matter, School of Science,
Xi'an Jiaotong University, Xi'an 710049, China}
\author{Fangfang Li}
\affiliation{MOE Key Laboratory for Nonequilibrium Synthesis and Modulation of Condensed Matter, School of Science,
Xi'an Jiaotong University, Xi'an 710049, China}
\author{Xing Wang}
\affiliation{MOE Key Laboratory for Nonequilibrium Synthesis and Modulation of Condensed Matter, School of Science,
Xi'an Jiaotong University, Xi'an 710049, China}
\author{Shuai Yin}
\affiliation{MOE Key Laboratory for Nonequilibrium Synthesis and Modulation of Condensed Matter, School of Science,
Xi'an Jiaotong University, Xi'an 710049, China}
\author{Jianhua Feng}
\affiliation{MOE Key Laboratory for Nonequilibrium Synthesis and Modulation of Condensed Matter, School of Science,
Xi'an Jiaotong University, Xi'an 710049, China}
\author{Xianming Zhou}
\affiliation{MOE Key Laboratory for Nonequilibrium Synthesis and Modulation of Condensed Matter, School of Science,
Xi'an Jiaotong University, Xi'an 710049, China}
\author{Yifang Gao}
\affiliation{MOE Key Laboratory for Nonequilibrium Synthesis and Modulation of Condensed Matter, School of Science,
Xi'an Jiaotong University, Xi'an 710049, China}
\author{Yuan Li}
\affiliation{MOE Key Laboratory for Nonequilibrium Synthesis and Modulation of Condensed Matter, School of Science,
Xi'an Jiaotong University, Xi'an 710049, China}
\author{Xiaohua Shi}
\affiliation{MOE Key Laboratory for Nonequilibrium Synthesis and Modulation of Condensed Matter, School of Science,
Xi'an Jiaotong University, Xi'an 710049, China}
\author{Jianxing Li}
\affiliation{MOE Key Laboratory for Nonequilibrium Synthesis and Modulation of Condensed Matter, School of Science, Xi'an Jiaotong University, Xi'an 710049, China}
\author{Xueguang Ren}
\affiliation{MOE Key Laboratory for Nonequilibrium Synthesis and Modulation of Condensed Matter, School of Science, Xi'an Jiaotong University, Xi'an 710049, China}
\author{Zhongfeng Xu}
\affiliation{MOE Key Laboratory for Nonequilibrium Synthesis and Modulation of Condensed Matter, School of Science, Xi'an Jiaotong University, Xi'an 710049, China}
\author{Zhigang Deng}
\affiliation{Science and Technology on Plasma Physics Laboratory, Laser Fusion Research Center, China Academy of Engineering Physics, Mianyang 621900, China}
\author{Wei Qi}
\affiliation{Science and Technology on Plasma Physics Laboratory, Laser Fusion Research Center, China Academy of Engineering Physics, Mianyang 621900, China}
\author{Shaoyi Wang}
\affiliation{Science and Technology on Plasma Physics Laboratory, Laser Fusion Research Center, China Academy of Engineering Physics, Mianyang 621900, China}
\author{Quanping Fan}
\affiliation{Science and Technology on Plasma Physics Laboratory, Laser Fusion Research Center, China Academy of Engineering Physics, Mianyang 621900, China}
\author{Bo Cui}
\affiliation{Science and Technology on Plasma Physics Laboratory, Laser Fusion Research Center, China Academy of Engineering Physics, Mianyang 621900, China}
\author{Weiwu Wang}
\affiliation{Science and Technology on Plasma Physics Laboratory, Laser Fusion Research Center, China Academy of Engineering Physics, Mianyang 621900, China}
\author{Zongqiang Yuan}
\affiliation{Science and Technology on Plasma Physics Laboratory, Laser Fusion Research Center, China Academy of Engineering Physics, Mianyang 621900, China}
\author{Jian Teng}
\affiliation{Science and Technology on Plasma Physics Laboratory, Laser Fusion Research Center, China Academy of Engineering Physics, Mianyang 621900, China}
\author{Yuchi Wu}
\affiliation{Science and Technology on Plasma Physics Laboratory, Laser Fusion Research Center, China Academy of Engineering Physics, Mianyang 621900, China}
\author{Zhurong Cao}
\affiliation{Science and Technology on Plasma Physics Laboratory, Laser Fusion Research Center, China Academy of Engineering Physics, Mianyang 621900, China}
\author{Zongqing Zhao}
\affiliation{Science and Technology on Plasma Physics Laboratory, Laser Fusion Research Center, China Academy of Engineering Physics, Mianyang 621900, China}
\author{Yuqiu Gu}
\affiliation{Science and Technology on Plasma Physics Laboratory, Laser Fusion Research Center, China Academy of Engineering Physics, Mianyang 621900, China}
\author{Leifeng Cao}
\affiliation{Advanced Materials Testing Technology Research Center, Shenzhen University of Technology, Shenzhen, 518118, China}
\author{Shaoping Zhu}
\affiliation{Science and Technology on Plasma Physics Laboratory, Laser Fusion Research Center, China Academy of Engineering Physics, Mianyang 621900, China}
\affiliation{Institute of Applied Physics and Computational Mathematics, Beijing 100094, China}
\affiliation{Graduate School, China Academy of Engineering Physics, Beijing 100088, China} 
\author{Rui Cheng}
\affiliation{Institute of Modern Physics, Chinese Academy of Sciences, Lanzhou 710049, China}
\author{Yu Lei}
\affiliation{Institute of Modern Physics, Chinese Academy of Sciences, Lanzhou 710049, China}
\author{Zhao Wang}
\affiliation{Institute of Modern Physics, Chinese Academy of Sciences, Lanzhou 710049, China}
\author{Zexian Zhou}
\affiliation{Institute of Modern Physics, Chinese Academy of Sciences, Lanzhou 710049, China}
\author{Guoqing Xiao}
\affiliation{Institute of Modern Physics, Chinese Academy of Sciences, Lanzhou 710049, China}
\author{Hongwei Zhao}
\affiliation{Institute of Modern Physics, Chinese Academy of Sciences, Lanzhou 710049, China}
\author{Dieter H.H. Hoffmann}
\affiliation{MOE Key Laboratory for Nonequilibrium Synthesis and Modulation of Condensed Matter, School of Science, Xi'an Jiaotong University, Xi'an 710049, China}
\author{Weimin Zhou} \email{zhouwm@caep.cn}
\affiliation{Science and Technology on Plasma Physics Laboratory, Laser Fusion Research Center, China Academy of Engineering Physics, Mianyang 621900, China}
\author{Yongtao Zhao} \email{zhaoyongtao@xjtu.edu.cn}
\affiliation{MOE Key Laboratory for Nonequilibrium Synthesis and Modulation of Condensed Matter, School of Science,
Xi'an Jiaotong University, Xi'an 710049, China}
\bibliographystyle{apsrev4-1}

\date{\today}

\begin{abstract}

The charge equilibration of laser-accelerated carbon ion beams in 2 mg/cm$^{3}$ foam target was investigated experimentally. The ions were generated through target normal sheath acceleration mechanism in laser-foil interaction scheme. This allows to get the equilibrium charge state in wide energy range near Bragg peak within a single shot. By using foam, the charge equilibration measurement in density regime between gas and solid state was firstly reached out experimentally.  It was found that the theoretical predictions with tabulated cross section data for gas target greatly underestimated the charge states. The experimental data are in close agreement with both semi-empirical formula as well as rate equation predictions based on ion-solid interactions. The important role of target density effects that increase the ionization probability and decrease the electron capture probability through frequent multi-collisions in foam are demonstrated. The double electron processes are shown to have little influence on the average charge states. The findings are essential for high energy density physics research where the foams are widely used, and have impacts on a broad range of applications in medical, biological and material fields. The method also provides a new approach to investigate the interaction mechanism of swift heavy ions in matter by taking advantage of the laser-accelerated short-pulse wide-energy range ions.
\end{abstract}

\maketitle
Since the invention of aerogel by S.S. Kistler in 1931 from Sodiumsilicat \cite{kistler1931coherent}, a tremendous development of material between gas and solid has taken place \cite{rosmej2005charge}. Today aerogel, jelly or foam targets are available. More than 99 percent of these materials are void, thus it is a highly porous material. Many applications are known e.g. storage of gases, as filter material, Cherenkov detector, or insulation, just to name a few. Especially, foam target was popularly used for enhanced charged particle generation and acceleration due to the high energy coupling efficiency with laser pulse \cite{rosmej2020high}. It was also successfully applied in recent experiment investigating the ion-matter interactions \cite{ren2023target}. Besides, the porous structure makes the charged particle transportation process more complex due to collective effects \cite{jiang2023branching}. Therefore, it is important to investigate the details of beam target interaction with this material. 

The charge exchange processes in interaction of ions with matter are the important subfield of atomic physics. They involve complex dynamic processes of collisional ionization and capture of electrons and influence the ion charge state distribution \cite{shima1989systematics,zhao2021benchmark}, the energy loss of ions or in another word the stopping power of the material \cite{betz1972charge,betz1983heavy,sharkov2016high,ziegler2013stopping,deutsch2016ion,gao2021projectile,2020Generation}. By this, the charge exchange processes are significant in the applications of swift heavy ions in material science, fusion energy as well as medical and biological fields. These processes also play important roles in heavy ion accelerator facilities, where gaseous or solid targets are used to generate highly charged ions to improve the acceleration efficiency of ion beams \cite{cheng2018warm,ren2017hydrodynamic,zhao2014high,zhao2016high,zhao2016high2}. A very well-known example is the stripper target at GSI which was placed between the Alvarez- and Wider\"{o}e section at 1.4 MeV/u \cite{erb1978gsi}.

In matter, the charge state stabilizes at an equilibrium value after a short distance, and from then starts to decrease with decreasing velocity of the ion. For solid state and gaseous targets \cite{shima1989systematics,nardi1982charge,Peter1991,scheidenberger1998charge,tolstikhina2018influence,fu2003atomic,wang2007separation}, a vast number of experimental data exist that constitutes a solid basis for theoretical models. The charge state evolution is well understood in principle and depends on the dynamic equilibrium of electron capture and ionization rates.  It was demonstrated that, in comparison with gasous target, the ions passing through solid target at similar line density show a markedly higher average charge state \cite{erb1978gsi}. This is known as the density effects caused by the high collision frequency that quenches the lifetime of excited states in solid targets as compared to gas targets \cite{shevelko2003target,shevelko2005target}. However, for the matter with density between gas and solid, very few data exist to pin down the charge state evolution behavior, especially in the energy range near Bragg peak. 

Here in this work, we performed the carbon ion charge equilibration measurement employing a tri-cellulose-acetate (TCA, C$_{9}$H$_{16}$O$_{8}$) foam with density of 2 mg/cm$^{3}$, which is about thousands of solid density. By using laser-accelerated carbon ions, the equilibrated charge states in wide-energy range of 5-12 MeV are obtained. The average equilibrium charge states were compared with various theoretical predictions. The importance of target density effects in this foam target are demonstrated. 

\begin{figure}
	\includegraphics[width=1\linewidth]{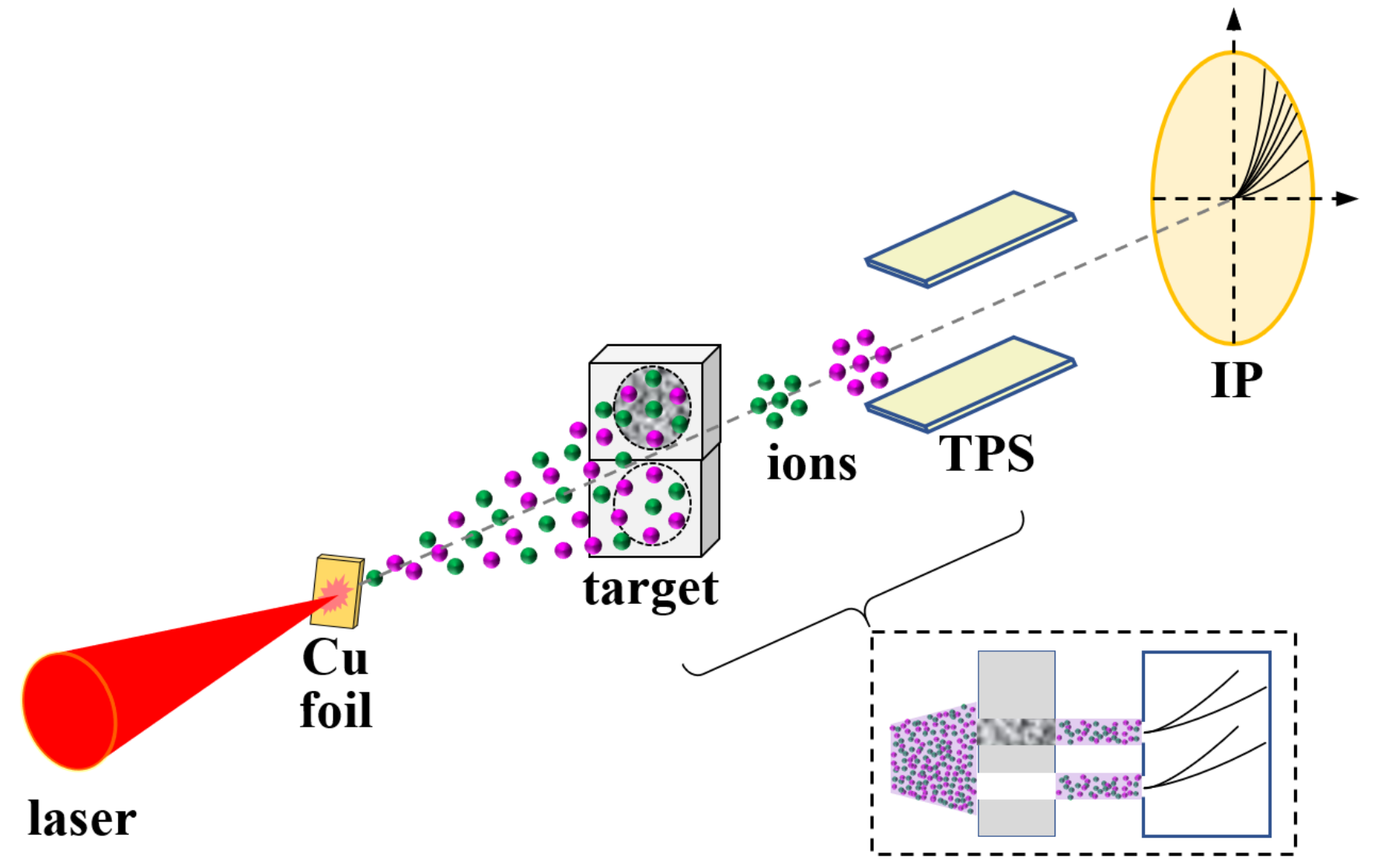}
	\caption{\label{fig1} Experimental set-up. A picosecond laser is focused onto a copper foil, generating an intense short-pulse of ions through TNSA mechanism. The energy spectra of carbon ions passing through the foam target and empty hole were measured with a dual-channel TPS coupled to the Fuji BAS-TR-type image plate (IP). }
\end{figure}

The experiment was carried out at the XG III laser facility of Laser Fusion Research Center in Mianyang, China. The schematic set-up of the experiment is shown in Fig. \ref{fig1}. A 10 µm copper foil was irradiated by a p-polarized laser pulse of 1053 nm central wavelength, 130 J energy, 10 $\mu$m focal spot and 843 ps duration at an incident angle of $10^{\circ}$ to generate short-pulse intense ion beam through the target normal sheath acceleration (TNSA) mechanism. The ion beam \cite{2013Charge,2015Coulomb,braenzel2018charge} consisted of protons and carbon ions with different charge states and a wide spectrum of kinetic energy. Here only carbon ions are discussed. 

The target is a tri-cellulose-acetate (TCA, C$_{9}$H$_{16}$O$_{8}$) foam with thickness of 1 mm in the direction of ion propagation. The density of the foam was fabricated through in-situ synthesis method as 2 mg/cm$^{3}$, which is about one thousandth of solid density. The density of the foam was benchmarked through measuring the energy loss of proton beam with radiative alpha sources.  The foam was widely used in series of experiments for laser acceleration, ion stopping \cite{ren2020observation,rosmej2011heating,rosmej2015hydrodynamic,ma2022plasma}, laboratory astrophysics \cite{ma2021laboratory} as well as charge exchange process in plasma \cite{ren2023target}. 

The laser-accelerated ions were directed onto the foam target and an empty channel (no foam case) symmetrically. In combination with the dual-channel Thompson Parabola Spectrometer (TPS), the energy spectra of carbon ions before and after interaction with the foam target can be obtained simultaneously in one laser shot. 

\begin{figure}
	\includegraphics[width=1\linewidth]{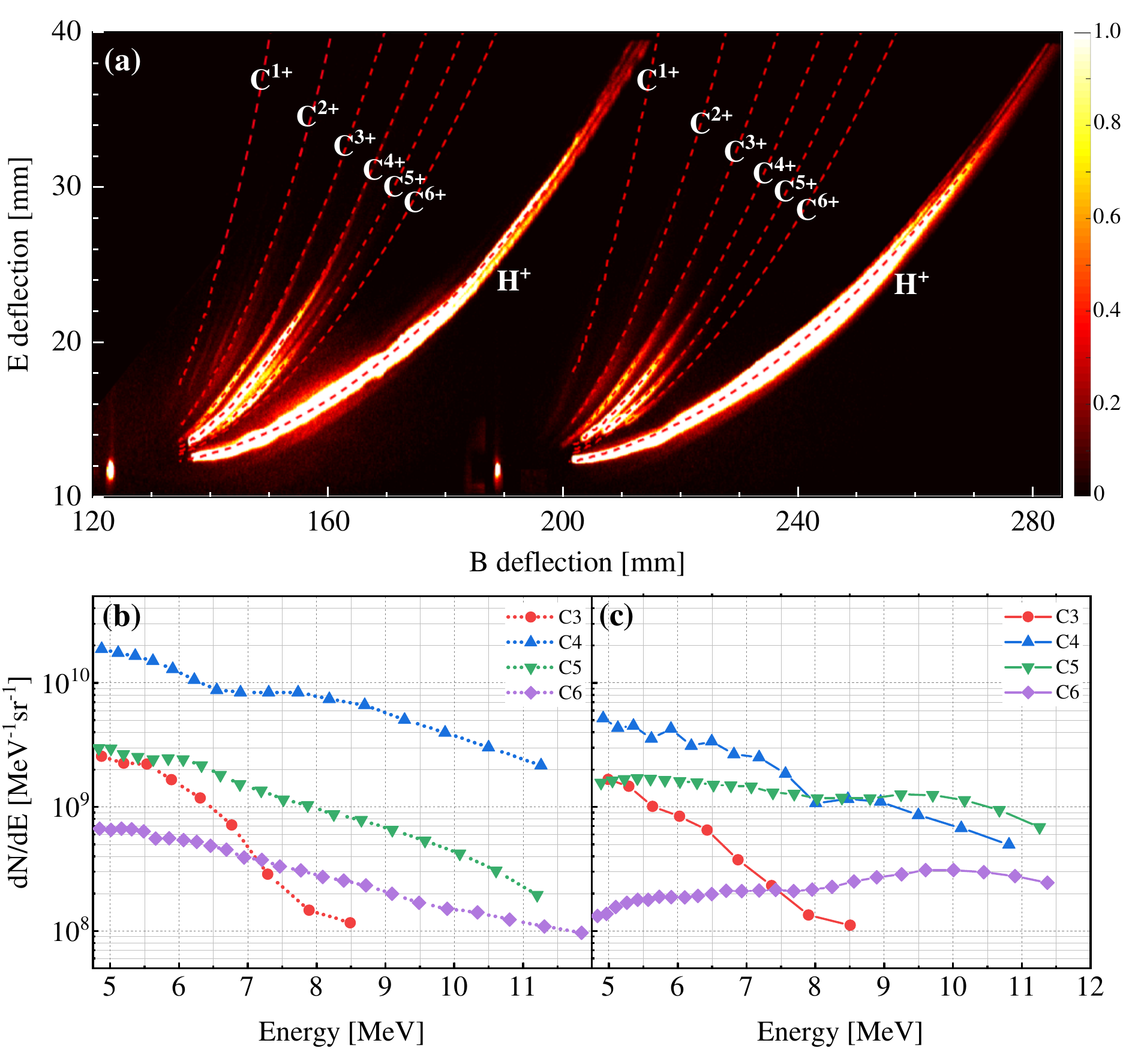}
	\caption{\label{fig2} (a) Typical deflection curves of laser-accelerated carbon ions without target (left) and passing through the foam (right). The converted energy spectra of carbon ions (b) without target and (c) passing through the foam.}
\end{figure}

The measured parabolic spectra of laser accelerated proton and carbon ions with and without foam were shown in Fig. \ref{fig2} (a) together with simulated deflection curves indicated by red dashed lines. The X and Y axis represented the deflection distances by magnetic field B and electrical field E of TPS, respectively. The zero points originate from the neutral particles like photons as well as hydrogen and carbon atoms et al.  It can be seen that the protons and carbon ions of different species are very well resolved, and the dominant carbon ion species are C$^{6+}$, C$^{5+}$, C$^{4+}$ and C$^{3+}$.

The deflection distances of the carbon ions for the cases without and with foam in Fig. \ref{fig2} (a) were converted to energies in range of 5 MeV to 11.5 MeV in Fig. \ref{fig2} (b) and (c). The energy spectra of the carbon ions were obviously changed after the ions passing through the foam due to both the charge transfer process as well as stopping process. Most notably, the number of C$^{4+}$ ions were significantly reduced with a drop of about 3.5 times from the maximum value. Another apparent change is with respect to the highly charged state ions C$^{5+}$ and C$^{6+}$, whose particle numbers increase significantly in the high energy range.

\begin{figure}
	\includegraphics[width=1\linewidth]{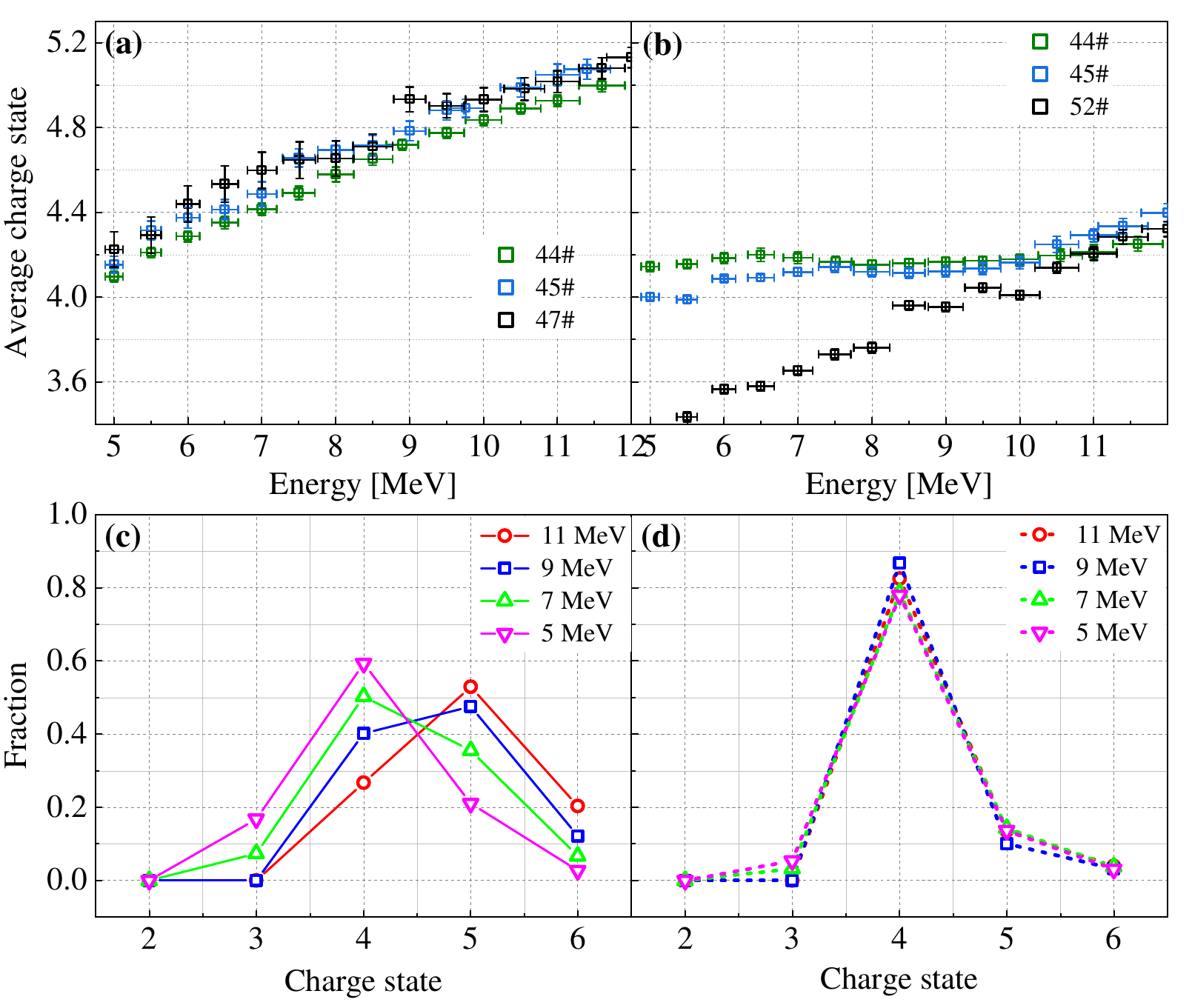}
	\caption{\label{fig3} Average charge states of carbon ions passing through foam (a) and without any target (b) for three shots. The error concerning energy and charge state originate the energy resolution of the TPS and statistical error, respectively. Charge state distribution evolution of carbon ions passing through foam (c) and without any target (d) versus kinetic energy.}
\end{figure}

According to LISE++ code \cite{tarasov2008lise++}, the equilibrium length of carbon ions in the foam target is less than 0.04 mm in the energy range of our experiment. That is about 25 times lower than the foam thickness. Hence the charge states of the outgoing carbon ions were actually in equilibrated distribution. Fig.\ref{fig3} (a) shows the equilibrated average charge states $\overline{Q}(E)=\sum_{q}qF_q (E) $, where $ F_q (E)$  is the fraction of carbon ions in charge state q and kinetic energy E, $\overline{Q}$ as a function of the kinetic energy for three shots. It can be seen that the average charge states of carbon ions passing through the foam can be very well repeated, because the charge states were in equilibrium and the equilibrated charge states depended only on the energy. This provides a way for laser-accelerated ion modulation. For comparison, the laser-accelerated carbon ions without target are presented in Fig. \ref{fig3} (b). The average charge state of the initial laser-accelerated carbon ions is less than 4.4 in the investigated energy range. As shown in Fig. \ref{fig3} (d), the dominant species are C$^{4+}$. However, after interacting with the foam target, the average charge states are enhanced. For example, for carbon ions at kinetic energy of 11 MeV, the average charge state is enhanced by about 17 \% from 4.2 to 4.9. Besides, after interacting with the foam target, the average charge state increased monotonously with the kinetic energy. That means with increasing energy, the fractions of highly ionized carbon ions (C$^{5+}$ and C$^{6+}$) are evidently increased. This were consistently shown in Fig. \ref{fig3} (c).  

\begin{figure}
	\includegraphics[width=1\linewidth]{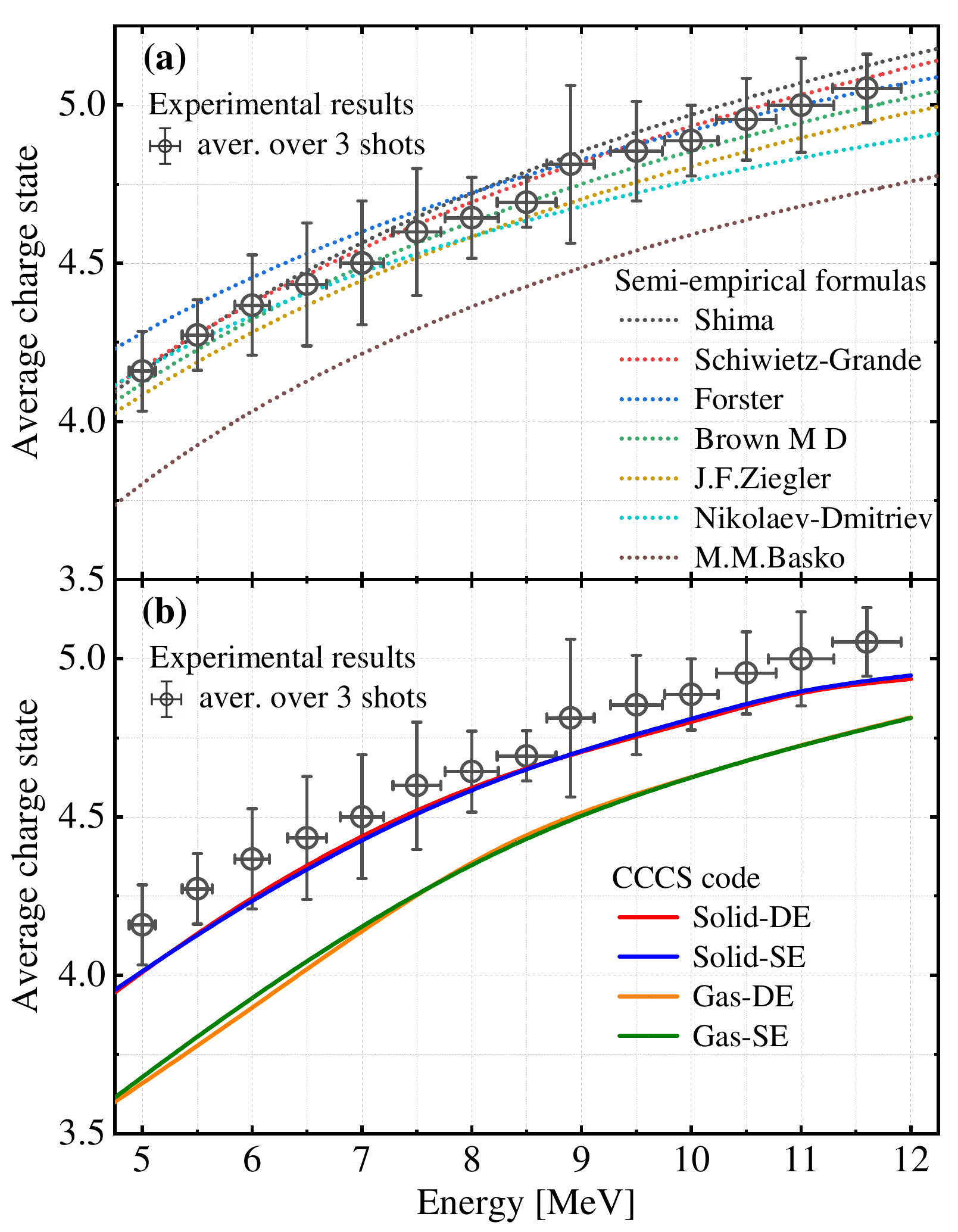}
	\caption{\label{fig4} The comparison of measured average charge states with theoretical predictions. The experimental data were averaged over three shots with errors representing the shot-to-shot fluctuations and statistical error. (a) The comparison with semi-empirical models. (b) The comparison with predictions through solving rate equations.}
\end{figure}

In Fig.\ref{fig4} (a), the measured average equilibrium charge states were compared to the commonly used semi-empirical formulas \cite{brown1972stopping,shima1982empirical,zeigler1985stopping,basko1983stopping,schiwietz2001improved,nikolaev1968equilibrium,forster1976stopping} that have been developed for ion interactions with solids based on the analytical expression of effective charge developed from Bohr’s stripping criterion \cite{bohr1960penetration} and Schiwietz’s fitting formula \cite{schiwietz2001improved}. Most of the models can well predicted the results, except for Basko's formula \cite{basko1983stopping}, which added the dependence of coulomb logarithm on Z and was more applicable for the particles with atomic number Z $\ge$ 20. Besides, the Nikolaev-Dmitriev model \cite{nikolaev1968equilibrium} underestimate our results at higher energy of 9-12MeV.

\begin{table*}\scriptsize
	\centering
	\caption{\label{Table 1} The charge transfer rates [$\times10^{12} s^{-1}$] for 6 MeV carbon ions with foam target.}
	\begin{adjustbox}{width=\textwidth,center}	
		\begin{tabular}{c c c c c}
			\toprule
			Charge state  & Single Electron Loss & Double Electron Loss & Single Electron Capture & Double Electron Capture \\
			\midrule
			C$^{0+}$ & 76.93 & 13.08 &       &       \\
			C$^{1+}$ & 73.04 & 12.42 & 0.27  &       \\
			C$^{2+}$ & 51.08 & 8.68  & 0.63  & 0.11  \\
			C$^{3+}$ & 12.27 & 2.09  & 1.51  & 0.26  \\
			C$^{4+}$ & 4.89  & 0.83  & 3.42  & 0.58  \\
			C$^{5+}$ & 0.46  &       & 6.74  & 1.15  \\
			C$^{6+}$ &       &       & 13.50 & 2.29  \\
			\bottomrule
	\end{tabular}
  \end{adjustbox}
\end{table*}

In Fig. \ref{fig4} (b), the measured average equilibrium charge states were compared to the rate equation predictions using tabulated cross sections from the Charge Changing Cross Sections (CCCS) code \cite{Novikov2021cccs,novikov2014methods} for gas and solid target considering single-electron (SE) and double electron (DE) transfer processes, respectively. When the cross sections of gas target were employed, the charge states were greatly underestimated. Once the cross sections of solid target were employed, the theoretical equilibrated charge states were obviously enhanced and approached our experimental data. The enhancement was attributed to the target density effect in dense matter which will on one hand increase the electron loss rate because of the two-step process of electron excitation and subsequent ionization in the secondary collision, and on the other hand decrease the electron capture rates because of the easy ionization of highly excited electrons that are captured from previous collision.

\begin{figure}
	\includegraphics[width=1\linewidth]{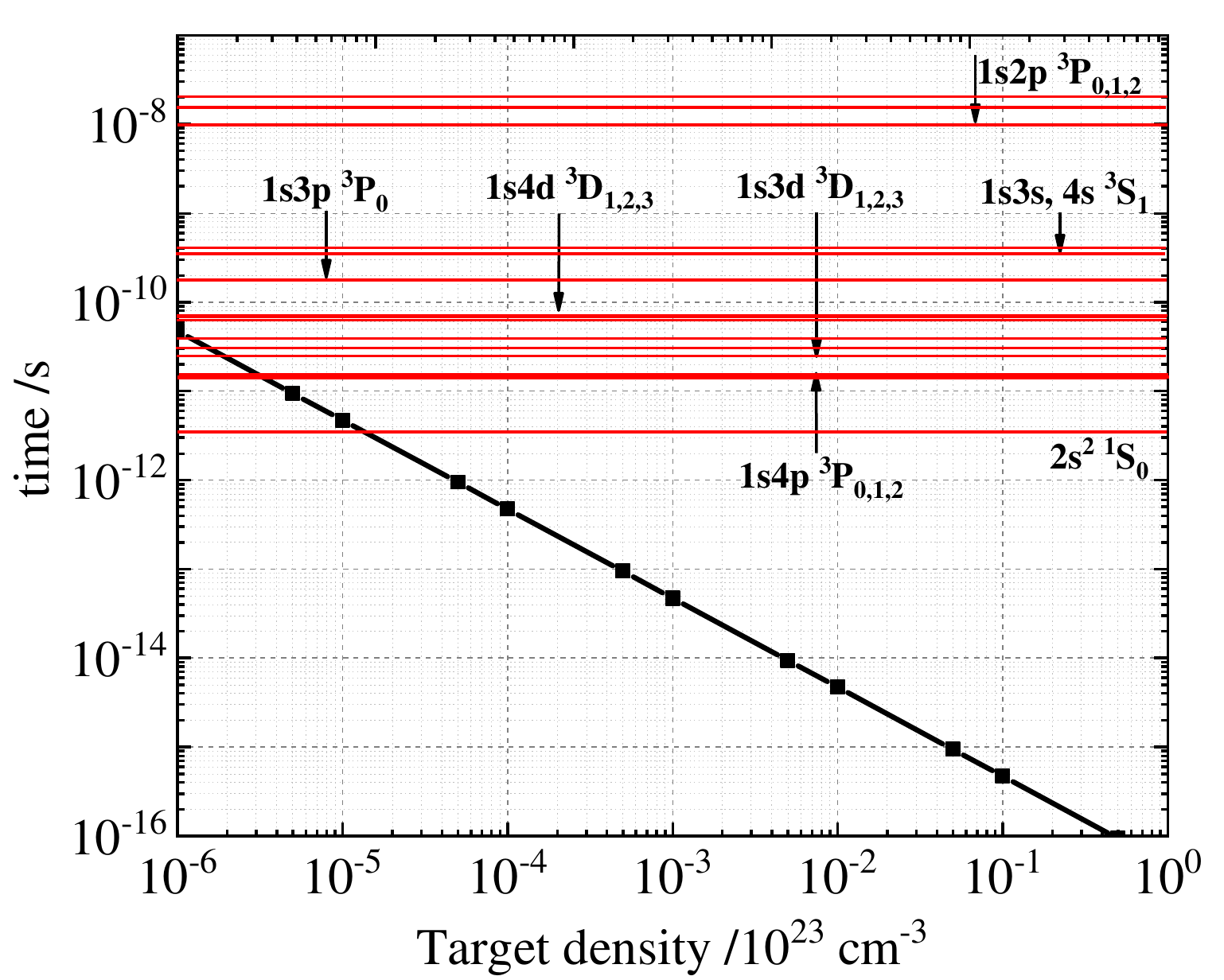}
	\caption{\label{fig5} The lifetime of the carbon excited states (red lines) as well as the collisional time-scale (black line) versus the target density.}
\end{figure}

The underlying fundamental physics of the target density effects is actually the balancing of the lifetime of excited states versus the collisional frequency. When the lifetime is larger than the collisional time-scale, the electron in excited states will be easily ionized before de-excitation, and the target density effects will play an important role. Fig. \ref{fig5} shows the comparison between the collisional time scale and the lifetime of some typical excited states of carbon ions as a function of target density. In the current collision system with foam target, the collisional time scale is much smaller than the lifetime of the excited states, in which case target density must be considered. This was demonstrated by the closer agreement of experimental data with theoretical predictions using solid-target cross sections, as shown in Fig.\ref{fig4} (b).

It was also shown in Fig. \ref{fig4} (b) that the double-electron transfer processes had negligible influence on the average charge state.  The charge transfer rates for the single-electron and double electron transfer processes of 6 MeV carbon atom and ions are listed in Table \ref{Table 1}. Both the single electron loss and electron capture rate are about 6 times higher than single electron loss and capture rates. This makes little difference on the charge state.

In summary, the equilibrated charge state distributions of laser-accelerated short-pulse carbon ions penetrating through the foam targets with energy near Bragg peak were measured. The average charge state of ions after the foam are well repeated. The stripping effect of foam on incident ions was clearly presented compared to the data of without any targets. The average charge states agree well with most of the semi-empirical formulas based on ion-solid interactions except for Basko's formula, which was more applicable for the particles with atomic number larger than 20. To deeply understand the charge transfer processes, the rate equations were solved with cross section data from CCCS code for both gas and solid targets. When cross sections for solid targets were employed, the charge states are obviously enhanced and are in closer agreement of our experimental results. That demonstrated the importance role of target density effects which can increased the ionization rates and decreased the electron capture rates. It was also shown that the double electron loss and capture processes have negligible influence on the charge states. This is the first time that the charge state measurement was performed in target with density between gas and solid. These results provide important data for high energy density physics, accelerator technologies as well as applications in medical, biological and material fields. In the current analysis, the cross sections for either gas or solid are employed. We hope our results will inspire more theoretical studies investigating the charge transfer process for states between gas and solid. Precise cross section measurements in those states are highly desirable.

The experiment was performed at the XG-III facility in Mianyang. The authors are grateful to the staff of Laser Fusion Research Center. The work was supported by National Key R\&D Program of China, No. 2019YFA0404900, and No. 2022YFA1603300, Chinese Science Challenge Project, No. TZ2016005, National Natural Science Foundation of China (Grant Nos. 12120101005, U2030104, 12175174, and 11975174), China Postdoctoral Science Foundation (Grant no. 2017M623145), State Key Laboratory Foundation of Laser Interaction with Matter (Nos. SKLLIM1807 and SKLLIM2106), and the Fundamental Research Funds for the Central Universities.

\bibliographystyle{unsrt}
\bibliography{ref}

\end{document}